\documentclass[twocolumn,showpacs,preprintnumbers,amsmath,amssymb]{revtex4}
\usepackage{graphicx}
\usepackage{dcolumn}
\usepackage{bm}
\begin{document}
\preprint{xxx}
\def\vk{\mbox{\boldmath $k$}}
\def\vr{\mbox{\boldmath $r$}}
\def\vv{\mbox{\boldmath $v$}}
\def\vs{\mbox{\boldmath $s$}}
\def\vz{\mbox{\boldmath $z$}}
\def\vM{\mbox{\boldmath $M$}}
\def\vome{\mbox{\boldmath $\omega$}}
\def\vOme{\mbox{\boldmath $\Omega$}}

\title{Dynamics of quantized vortices in superfluid helium
and rotating Bose-Einstein condensates}
\author{Makoto Tsubota, Kenichi Kasamatsu and Tsunehiko Araki}
\affiliation{Department of Physics, Osaka City University, \\Sumiyoshi-ku,
Osaka 558-8585, Japan}
\date{\today}

\newpage
\begin{abstract}
In this article, we review the research on the dynamics of quantized
vortices in superfluid helium and rotating Bose-Einstein condensates with
emphasis on the recent research done by our group.
A quantized vortex is a topological defect that arises from the order
parameter in Bose-Einstein condensation in which frictionless superfluid
flows with quantized circulation around each vortex.

A quantized vortex was both predicted and discovered first in superfluid
$^4$He which was the first example of a Bose-Einstein condensate. Quantized
vortices have been thoroughly studied in superfluid $^4$He; one of the
principal problems was the superfluid turbulent state consisting of a tangle
of quantized vortices in thermal counterflow.  More recently, the interest
has shifted to the nature of superfluid turbulence, apart from the case of
counterflow. After briefly reviewing the earlier research and describing the
current problems, we focus our review on superfluid turbulence and vortex
filament dynamics.  One of the important problems is how superfluid
turbulence relates to classical turbulence.  Superfluid turbulence was
recently shown to have an energy spectrum consistent with the Kolmogorov
law, which is an important statistical law in fully developed classical
turbulence. We also describe the diffusion of an inhomogeneous vortex tangle
with relation to the observed decay of vortices at very low temperatures
where the normal fluid component is so negligible that the usual mutual
friction does not work as a decay mechanism. In connection to the above
discussion of groups of vortices, we describe the vortex states that appear
in a rotating channel with counterflow. Rotational effects cause the
vortices to form ordered arrays, whereas counterflow effects tend to cause
disordered vortex tangles; the competition of these two effects makes a new
state of ``a polarized vortex tangle" which opens up superfluid phase
diagrams as a new area of study.

In the specific field of atomic-gas Bose-Einstein condensation, we discuss
recent numerical analysis of the Gross-Pitaevskii equation that describes
the structure and dynamics of the order parameter.  Consistent with
observations, the simulated condensate starts an elliptic oscillation after
the rotation is turned on, which induces the surface-mode excitations. The
vortices develop from these surface excitations and then enter the bulk
condensate, eventually forming a vortex lattice. When a condensate is held
in a quadratic-plus-quartic combined potential, the fast rotation makes ``a
giant vortex" in which most vortices are absorbed into a central hole around
which the superflow circulates.
\end{abstract}
\maketitle
\newpage

\section{INTRODUCTION}
In Bose-Einstein condensations(BECs), a macroscopic fraction of bosons
constitute a condensate wave function from which quantized vortices can
form.  A quantized vortex is a topological defect characteristic of a
Bose-Einstein condensate that has a quantized circulation of superflow and a
thin vortex core of the order of the microscopic coherence length.

Quantized vortices were first predicted theoretically by Onsager
\cite{onsager} and Feynman \cite{Feynman}. Since their discovery in superflu
id $^4$He \cite{vinen2}, research on quantized vortices has become one of
the main subjects in low temperature physics \cite{donnelly}. On the other
hand, the achievement of Bose-Einstein condensation in dilute atomic gases
\cite{ps} in 1995 has opened up a new research field, in which quantized
vortices can be well controlled and visualized using new optical techniques.
The study of quantized vortices has an important role in understanding the
physics in both the traditional field of superfluid helium and the novel
field of atomic-gas Bose-Einstein condensates (BECs). Being motivated by the
recent developments in these fields, our group has studied theoretically and
numerically the dynamics of quantized vortices in superfluid $^4$He and
atomic BECs. The purpose of this article is to review the physics of
quantized vortices in both systems with emphasis on the recent activities in
our group \cite{movie}.

Section \ref{question} summarizes the properties of quantized vortices and
their differences from  vortices in a classical fluid. Section \ref{helium}
is concerned with the recent research of superfluid turbulence. Most of the
early studies of superfluid turbulence were limited to the case of thermal
counterflow, which has no analogy with classical turbulence. However, we
also discuss recent experimental studies that help clarify the similarity
between superfluid turbulence and classical turbulence. After describing the
equation of motion of quantized vortex filaments, we discuss the energy
spectrum of a vortex tangle in superfluid turbulence, the diffusion of an
inhomogeneous vortex tangle, and a vortex tangle polarized by rotating the
vessel. The vortex dynamics and vortex lattice formation of rotating BECs
are described in Sec.\ref{roteBEC}. Here we discuss the theoretical and
numerical analysis of the problem. Recent experimental and numerical results
show that elliptic deformation of the condensate and surface-mode
excitations cause a vortex lattice to form. Section \ref{concle} is devoted
to conclusions.

\section{WHAT IS A QUANTIZED VORTEX?}\label{question}
Below a critical temperature in an ideal Bose gas, a finite fraction of the
particles occupies the same single-particle ground state and forms a BEC.
When the particles have mutual interaction, single-particle states are no
longer meaningful. However  a condensate wave function $\Psi(\vr,t)$ is
still  defined as the ensemble average of the quantum amplitude for removing
a particle at position $\vr$  from the condensate. When a BEC is held in an
external potential $V(\vr)$, the dynamics of $\Psi(\vr,t)$ is described by
the Gross-Pitaevskii(GP) equation
\begin{equation}
i \hbar \frac{\partial \Psi(\vr,t)}{\partial t} = \biggl( - \frac{
\hbar ^2}{2m}\nabla^2
+ V(\vr) +g |\Psi(\vr,t)|^{2} \biggr) \Psi(\vr,t),
\label{gpeq}
\end{equation}
where $g=4\pi \hbar^2 m/a$ represents the strength of interaction
characterized by the s-wave scattering length $a $ and $m$ is the mass of
each particle. Writing $\Psi = | \Psi | \exp (i \theta)$, the squared
amplitude $|\Psi|^2$ is the condensate density and the gradient of the phase
$\theta$ gives the superfluid velocity  $\vv_s = (\hbar/m) \nabla \theta$,
which is a frictionless flow of the condensate. As a result, the circulation
of $\vv_s$ around a closed path ${\cal C}$ in the fluid is quantized as
\begin{equation}
\oint_{\cal C} d\vs \cdot \vv_s = \frac{\hbar}{m} \oint_{\cal C} d\vs \cdot
\nabla \theta
= n \kappa  \quad (n=0, \pm 1, \pm 2, \ldots),
\end{equation}
with the quantum of circulation $\kappa =h/m$. Such a vortex with a
quantized circulation is called a quantized vortex; it occurs in superfluid
$^4$He, superfluid $^3$He, and atomic BECs.

A quantized vortex is different from a vortex in a classical viscous fluid.
First, the circulation is quantized, which is contrary to the classical
vortex that can have any value of circulation. Second, a quantized vortex is
a vortex of inviscid superflow. Thus, it cannot decay by the viscous
diffusion of vorticity that occurs in the classical system. A quantized
vortex can decay by shortening the length of the core through
mutual friction with the normal fluid, by breaking into smaller and smaller
vortex rings through reconnections and finally changing to some elementary
excitations, and by transferring energy to smaller length scales through a
Kelvin wave cascade, followed by acoustic emission. Third, the core of a
quantized vortex is very thin, being the order of the coherence length
defined by $\xi=\hbar/(\sqrt{2mg}| \Psi |)$, which is submicron in an
atomic-gas BEC and only a few angstroms in superfluid $^4$He. Because the
vortex core is very thin and does not decay by diffusion, it is easy
identify the position of a quantized vortex in the fluid. These properties
make a quantized vortex more stable and definite than a classical vortex.

\section{DYNAMICS OF QUANTIZED VORTICES IN SUPERFLUID HELIUM}\label{helium}
\subsection{EARLY RESEARCH IN SUPERFLUID TURBULENCE AND VORTEX DYNAMICS}
Below the $\lambda$ temperature at about 2.2 K at vapor pressure, liquid
$^4$He enters the superfluid state called helium II. Although the condensate
is depressed by interatomic interactions, this transition is microscopically
caused by Bose-Einstein condensation. Helium II behaves like an irrotational
ideal fluid, and its characteristic phenomena can be explained well by the
two-fluid model \cite{twofluid}. The two fluid model states that superfluid
$^4$He is a mixture of inviscid superfluid and viscous normal fluid with the
mixing ratio depending on temperature; the superfluid corresponds to the BEC
\cite{kobayashi} and the normal fluid has all of the thermal excitations of
the system.

Most early experimental studies focused on thermal counterflow. When helium
II is confined in a channel closed at one end with a heater, superfluid
enters from the open end and flows toward the heater. On reaching the
heater, normal fluid is created, which then flows back toward the open end.
This situation is called thermal counterflow. It is laminar at low relative
velocities, but when the relative velocity exceeds some critical velocity,
the superflow becomes turbulent and has dissipation. The concept of
superfluid turbulence was introduced by Feynman \cite{Feynman} who proposed
that the superfluid turbulent state consists of a disordered set of
quantized vortices, which is called a vortex tangle, that dissipates via
mutual friction between the vortex cores and the normal flow. The mutual
friction was introduced by Gorter and Mellink during their study of heat
conduction through helium II in which they measured the relationship between
temperature gradient and heat-current density \cite{gorter}. Later,
independently of Feynman's proposal,  Hall and Vinen started the pioneering
experimental works by making the simultaneous measurement of the temperature
gradient and the second sound attenuation across counterflow.  They
discovered extra second sound attenuation and understood it in terms of the
two fluid model with a generalized mutual friction \cite{hallvinen1}. After
Feynman's proposal, Hall and Vinen confirmed that mutual friction occurs by
observing the anisotropy of second sound attenuation in uniformly rotating
helium II \cite{hallvinen2}. Later, Vinen investigated in detail the
temperature gradient and the second sound attenuation across thermal
counterflow to understand how the mutual friction changed when the heat
current was suddenly changed. He found that the mutual friction came from
rotational motion of superfluid, eventually establishing the picture of
superfluid turbulent state consistent with Feynman's picture \cite{vinen1}.
The quantum of circulation was also measured by Vinen when he investigated
the oscillation of a vibrating wire in rotating helium II \cite{vinen2}.
There were many experimental studies of superfluid turbulence after Vinen's
papers \cite{tough}, chiefly on thermal counterflow.

The nonlinear and nonlocal dynamics of vortices long delayed progress in
further microscopic understanding of the vortex tangle. But Schwarz overcame
these difficulties \cite{Schwarz}. His most important contribution was his
direct numerical simulation of vortex dynamics connected with dynamical
scaling analysis, thus enabled others to calculate such physical quantities
as the vortex line density, various anisotropic parameters, and the mutual
friction force \cite{Schwarz}. The observable quantities obtained with
Schwarz's theory agreed well with the experimental results of the steady
state of the vortex tangle. This research field pioneered by Schwarz has
generated many new areas of study in vortex dynamics, such as the flow
properties in channels \cite{Samuels92,Barenghi,Aarts,Penz}, sideband
instability of Kelvin waves \cite{Samuels90},  vortex array in rotating
superfluid\cite{Tsubota95}, and vortex pinning \cite{Tsubota93,Tsubota94}.

\subsection{RECENT STUDIES ON SUPERFLUID TURBULENCE}
Although our understanding of superfluid turbulence and vortex dynamics has
made significant gains, the relation between superfluid turbulence and
classical turbulence remains a major unsolved problem \cite{vinen4,vinen00}.
Probably, this situation has arisen because most studies of superfluid
turbulence have been devoted to thermal counterflow which has no analogy
with classical turbulence.

We now discuss several relevant features of classical turbulence. In
particular, we describe the finding that superfluid turbulence can give a
clearer example of the inertial range than does classical turbulence. It has
been known for over a hundred years that a classical viscous fluid becomes
turbulent when the Reynolds number exceeds about 1000 \cite{frisch}. The
Reynolds number is given by the ratio of the nonlinear inertial term to the
viscous term in the Navier-Stokes equation; therefore, a large Reynolds
number means that viscosity is less important compared with the inertial
term. Apart from some boundary effects, at high Reynolds numbers,
fully-developed, homogeneous turbulence follows some universal statistical
laws. It turns out to be more helpful to consider energy and velocity in
wave number $k$ space based on the spatial Fourier transform $\vv(\vk)$ of
the velocity field $\vv(\vr)$. The important statistical law is represented
in the energy spectrum $E(k)$, where $k=|\vk|$ characterizes a length scale
$k^{-1}$ and $E(k) dk$ is the average turbulent energy per unit mass in the
range of the wave numbers $dk$. As an example, consider stationary grid
turbulence in which a flow passing through a grid with the mesh size $D$ is
close to being homogeneously turbulent far behind the grid. The behavior of
$E(k)$ is classified into three ranges depending on the wave number. In the
first energy-containing range, the energy is injected into the velocity
field at the wave number $k \sim D^{-1}$, by a rate $\epsilon$ per unit
mass. This energy flows into the second inertial range with higher wave
number; this inertial range is characteristic of fully developed turbulence.
Here the energy, not being dissipated, just flows toward the higher $k$
region by the energy flux $\epsilon$. The inertial range has no
characteristic scale because it is sustained by the Richardson cascade
process in which large eddies are continuously broken up self-similarly to
smaller ones by a nonlinear interaction. Since the inertial range should be
characterized only by the energy flux $\epsilon$ and the wave number $k$,
the energy spectrum follows the well known Kolmogorov law
\begin{equation}
E(k) = C_K \epsilon^{2/3} k^{-5/3}.
\end{equation}
Here, the Kolmogorov constant $C_K $ is found to be about 1.5
experimentally. The energy flowing from the inertial range into the third
energy-dissipative range is finally dissipated by the dissipation rate
$\epsilon$ at about the Kolmogorov wave number $k_K=(\epsilon/\nu^3)^{1/4}$
with the kinematic viscosity $\nu$. The important point is that the inertial
range has a universal spectrum that is independent of the specific
dissipative mechanism that operates in the energy-dissipative range. It is
for this reason that superfluid turbulence is expected to follow the
Kolmogorov law.

Recent experimental research studied superfluid turbulence not in thermal
counterflow, thus finding support for the Kolmogorov law.  Maurer and
Tabeling measured local pressure fluctuations in helium flows driven by two
counter-rotating disks in a range of temperature between 1.4 and 2.3 K and
obtained the Kolmogorov spectrum above and below $T_{\lambda}$
\cite{maurer}. A group at the University of Oregon reported in a series of
papers \cite{smith,stalp1,skrbek,stalp2} the observed attenuation of second
sound behind a grid that moved steadily through helium II at temperatures
above 1 K. Among these works, Stalp {\it et al.} measured the decay of grid
turbulence in helium II and showed that the experimental results were
consistent with a classical model of energy spectrum that included the
Kolmogorov law \cite{stalp1}. Then, Vinen analyzed the similarity between
superfluid turbulence and classical turbulence \cite{vinen00} and showed the
importance of length scales for understanding the energy of the velocity
field. For example, although superfluid turbulence is made of a tangle of
quantized vortices, the situation depends on whether the length scale is
larger or smaller than the vortex line spacing $\ell$. Furthermore, at
relatively high temperatures, the normal fluid and the superfluid are
coupled by mutual friction at all relevant scales larger than $\ell$, and
hence the system follows the Kolmogorov law at these scales. This theory was
shown to be consistent with the previous experimental results by Stalp {\it
et al.}, thus forming a consistent picture of superfluid turbulence with an
appreciable component of normal fluid.

Then appears a very important question; how is the energy spectrum of
superfluid turbulence at very low temperatures where the normal fluid
component is negligible? This is significant in the following reasons.
First, superflow guarantees that any dissipation could work only at some
very small scale because of its inviscosity. This means the presence of the
definite inertial range in which the energy is transferred from large scale
to smaller scales by the Richardson cascade process. The resulting small
vortices whose size becomes close to the coherence length would be unable to
keep its vortex nature to change into some elementary excitations, or
dissipate by acoustic emission \cite{vinen00}, but the inertial range should
be independent of such dissipative mechanisms.  Second, superfluid
turbulence is made of a tangle of quantized vortices. The inertial range of
a classical fluid is believed to be sustained by the Richardson cascade,
while the identification of each vortex in a turbulence is rather obscure.
On the other hand,  quantized vortices are definite and stable as described
in Sec. II,  so that the physical picture of the inertial range and the
Richardson cascade could be much clearer than that in a classical
turbulence. Therefore superfluid turbulence without the normal fluid
component may give a typical and simple prototype of turbulence.

There are no experimental studies of the energy spectrum of superfluid
turbulence at very low temperatures. Nevertheless, the two numerical works
on this topic indicate that the Kolmogorov law also applies to superfluid
turbulence at very low temperatures. Starting from a flow that mimicked a
Taylor-Green vortex, Nore {\it et al.} studied the decaying turbulence by
using the GP equation\cite{nore}. The energy spectrum succeeded in showing a
transient Kolmogorov form over a range of wave numbers less than
$\ell^{-1}$, but the acoustic emission is closely connected with the vortex
dynamics and the situation is complicated.  Our group studied the energy
spectrum under the vortex filament formulation and found that the Kolmogorov
law should apply to superfluids at very low temperatures \cite{araki}.  This
formulation and the results will be described  in the following subsections.

\subsection{DYNAMICS OF QUANTIZED VORTEX FILAMENTS}
  All numerical calculations in this chapter were done under the vortex
filament formulation. This subsection describes the equations of motion of
quantized vortex filaments and the method of
the numerical simulation.  A vortex filament is an idealized model of
rotational flow in which the vorticity is confined to a small core region
around a one-dimensional line embedded in the three-dimensional flow
\cite{samuels00}. In the field of classical fluid dynamics, vortex filaments
have been a useful model since the early 1980s for understanding the
geometry and dynamics of a flow. However, in very recent years interest in
this model has decreased because it is not capable of describing either the
complicated turbulent structures of real systems or solutions of the
Navier-Stokes equation \cite{frisch}. But contrary to the situation in a
classical fluid, vortex filaments are realistic in superfluids such as
helium II,  because here the cores are very thin and relatively stable.  A
quantized vortex filament may be likened to a "skeleton" of a vortex because
the quantum mechanical constraint means that each vortex has the ground
state circulation and a simple potential flow around its core. Our present
main interest is to understand if a tangle of such quantized vortex
filaments can still create the statistical law such as the Kolmogorov
spectrum. This subsection is devoted to the description of the vortex
filament formulation \cite{Schwarz,tsubota00}, followed by a discussion of
the energy spectrum in the next subsection.

The vortex filament formulation represents a quantized vortex as a filament
passing through the fluid and having a definite direction corresponding to
its vorticity. Except for the thin core region, the superflow velocity field
has a classically well-defined meaning and can be described by ideal fluid
dynamics. The velocity at a point $\vr$ due to a filament is given by the
Biot-Savart expression
\begin{equation}
\vv_{s} (\vr )=\frac{\kappa}{4\pi}\int_{\cal L} \frac{(\vs_1 - \vr) \times
d\vs_1}{|\vs_1-\vr|^3},
\label{BS}
\end{equation}
where $\kappa$ is the quantum of circulation. The filament is represented by
the parametric form $\vs = \vs(\xi, t)$ with the one-dimensional coordinate
$\xi$ along the filament. The vector $\vs_1$ refers to a point on the
filament and the integration is taken along the filament. Helmholtz's
theorem for a perfect fluid states that the vortex moves with the superfluid
velocity. Attempting to calculate the velocity $\vv_{s}$ at a point
$\vr=\vs$ on the filament makes the integral diverge as $\vs_1 \rightarrow
\vs$. To avoid this divergence, we separate the velocity $\dot{\vs}$ of the
filament at the point $\vs$ into two components \cite{Schwarz}:
\begin{equation}
\dot{\vs} =\frac{\kappa}{4\pi}\vs' \times \vs'' \ln \left(
\frac{2(\ell_+ \ell_-)^{1/2}}
{e^{1/4} a_0}\right) + \frac{\kappa}{4\pi}\int_{\cal L}^{'} \frac{(\vs_1 -
\vr)
\times d\vs_1}{|\vs_1-\vr|^3}.
\label{sdot}
\end{equation}
The first term is the localized induction field arising from a curved line
element acting on itself, and $\ell_+$ and $\ell_-$ are the lengths of the
two adjacent line elements, after discritization, that hold the point $\vs$
between them. The prime denotes differentiation with respect to the arc
length $\xi$. The mutually perpendicular vectors $\vs'$, $\vs''$ and $\vs'
\times \vs''$ point along the tangent, the principal normal and the binormal
at the point $\vs$, respectively, and their magnitudes are 1, $R^{-1}$, and
$R^{-1}$ with the local radius $R$ of curvature. The parameter $a_0$ is a
cutoff parameter equal to the core radius. Thus, the first term represents
the tendency for the points to move the local point $\vs$ along the binormal
direction with a velocity inversely proportional to $R$. The second term
represents the nonlocal field obtained by integrating the integral of Eq.
(\ref{BS}) along the rest of the filament.

The approximation that neglects the nonlocal terms and replacs Eq.
(\ref{sdot}) by $\dot{\vs} = \beta \vs' \times \vs''$ is called the
localized induction approximation (LIA). Here the coefficient $\beta$ is
defined by $\beta = (\kappa/4\pi) \ln \left( c \langle R \rangle/a_0
\right)$, where $c$  is a constant of order 1 and $(\ell_+\ell_-)^{1/2}$ is
replaced by the mean radius of curvature $\langle R \rangle$ along the
length of the filament.  Most of Schwarz's numerical studies on vortex
tangles used the LIA because this approximation can greatly reduce
computation times. Although the method is effective for the analysis of
dense tangles (due to cancellations between nonlocal contributions), it does
not include the intervortex interaction properly. Because our present
problems described in the following subsections need to take much account of
the interaction, our numerical simulations are done by not the LIA but the
fully Biot-Savart law of Eq. (\ref{sdot}) .

  A better understanding of vortices in a real system results when one
includes the boundaries in the analyses. For this, the boundary-induced
velocity field $\vv_{s,b}$ is added to $\vv_{s}$ so that the superflow can
satisfy the boundary condition of an inviscid flow. If the boundaries are
specular plane surfaces, $\vv_{s,b}$ is just the field due to an image
vortex made by reflecting the vortex into the plane and reversing its
direction of vorticity. To allow for another, presently unspecified, applied
field, we include $\vv_{s,a}$. Hence, the total velocity $\dot{\vs}_0$ of
the vortex filament without dissipation is
\begin{eqnarray}
\dot{\vs}_0 &=& \frac{\kappa}{4\pi}\vs' \times \vs'' \ln \left( \frac{2(\ell_+
\ell_-)^{1/2}}{e^{1/4} a_0}\right)  \nonumber \\ 
& & + \frac{\kappa}{4\pi}\int_{\cal L}^{'}
\frac{(\vs_1 - \vr) \times d\vs_1}{|\vs_1-\vr|^3}  
+\vv_{s,b}(\vs)+\vv_{s,a}(\vs).
\label{s0dot}
\end{eqnarray}
\noindent
At finite temperatures the mutual friction between the vortex core and the
normal flow $\vv_n$ is important. Including this term, the velocity of $\vs$
is given by
\begin{equation}
\dot{\vs} =\dot{\vs}_0 + \alpha \vs' \times (\vv_n - \dot{\vs}_0)
- \alpha' \vs' \times [\vs' \times (\vv_n - \dot{\vs}_0)],
\label{sdotmf}
\end{equation}
where $\alpha$ and $\alpha'$ are the temperature-dependent friction
coefficients, and $\dot{\vs}_0$ is calculated from Eq.(\ref{s0dot}).

The method used in the numerical simulation is similar to that of Schwarz
\cite{Schwarz} and described in detail in our paper \cite{tsubota00}. A
vortex filament is represented by a single string of points with a distance
$\Delta\xi$. The vortex configuration at a given time determines the
velocity field in the fluid, thus moving the vortex filaments according to
Eqs. (\ref{s0dot}) and (\ref{sdotmf}). Both local and nonlocal terms are
represented by means of line elements connecting two adjacent points. As the
vortex configuration develops and, particularly, when two vortices approach
each other, the distance between neighboring points can change. Then it is
necessary to add or remove points properly to retain sufficient local
resolution. Through the cascade process, a large vortex can break up through
many reconnections, eventually becoming a vortex that is smaller than the
space resolution $\Delta \xi$. The simulations cannot follow the dynamics
beyond this point, so these vortices are eliminated numerically.

It is important to properly include vortex reconnection when simulating
vortex dynamics.  A numerical study of a classical fluid showed that the
close interaction of two vortices leads to their reconnection, chiefly
because of the viscous diffusion of the vorticity \cite{Boratav}. Schwarz
assumed that  two vortex filaments reconnect when they get close within a
critical distance, and showed that the statistical quantities such as vortex
line density were not sensitive to how to make reconnections.  Even after
the Schwarz's works, it was still unclear whether quantized vortices can
actually reconnect or not. However Koplik and Levine solved directly the GP
equation to show the two close quantized vortices reconnected even in an
inviscid superfluid \cite{Koplik}. More recent simulations showed that
reconnections were accompanied by acoustic emissions
\cite{leadbeater,ogawa}. As the vortex filament formulation cannot represent
the reconnection process itself, we assume that two vortices reconnect when
they approach within the space resolution
$\Delta \xi$. The detail of this procedure is discussed in our paper
\cite{tsubota00}.

\subsection{ENERGY SPECTRUM OF SUPERFLUID TURBULENCE}
  To understand the energy spectrum and determine whether or not superfluids
follow the Kolmogorov law at very low temperatures, one must calculate the
energy spectrum. The energy spectrum is originally calculated by the Fourier
transform of the fluid velocity $\vv(\vr)$. The superfluid velocity
$\vv_s(\vr)$ is determined by the configuration of quantized vortices in our
vortex filament formulation. Therefore, one can calculate the energy
spectrum directly from the configuration of vortices. This is a big
advantage of this formulation because it makes the numerical calculation of
the spectrum less time-consuming than the method via  $\vv(\vr)$. Using the
Fourier transform $\vv_s(\vk)=(2\pi)^{-3} \int d\vr \vv_s(\vr)\exp (-i \vk
\cdot \vr) $ and Parseval's theorem $\int d\vk |\vv_s(\vk)|^2=(2\pi)^{-3}
\int d\vr |\vv_s(\vr)|^2$, the kinetic energy of the superfluid velocity per
unit mass is
 \begin{equation}
E=\frac12 \int d\vr |\vv_s(\vr)|^2=\frac{(2\pi)^3}{2} \int d\vk |\vv_s(\vk)|^2.
\end{equation}
The vorticity $\vome(\vr)={\rm rot} \vv_s(\vr)$ is represented in Fourier
space as $\vv_s(\vk)=i \vk \times \vome(\vk) /|\vk|^2$, so that we have
$E=\left( (2\pi)^3/2 \right) \int d\vk |\vome(\vk)|^2/|\vk|^2$. The
vorticity $\vome(\vr)=\kappa \int d\xi \vs'(\xi) \delta (\vs(\xi)-\vr$) in
the vortex filament formulation is rewritten as $\vome(\vk)=\left(
\kappa/(2\pi)^2 \right) \int d\xi \vs'(\xi) \exp (-i \vs(\xi) \cdot\vk)$.
Using the definition of the energy spectrum $E(k)$ from $E= \int_0^{\infty}
dk E(k)$, these relations yield
\begin{eqnarray}
E(k)=\frac{\kappa^2}{2(2\pi)^3} \int \frac{d\Omega_k}{|\vk|^2} \int\int d\xi_1
d\xi_2 \vs'(\xi_1)\cdot \vs'(\xi_2) \nonumber \\
\times \exp (-i \vk \cdot(\vs(\xi_1)-\vs(\xi_2))),
\label{spectrum}
\end{eqnarray}
where $d\Omega_k=k^2 \sin \theta_k d\theta_k d\phi_k$ is the volume element
in spherical coordinates. This formula connects the energy spectrum directly
with the vortex configuration.

\begin{figure}[btp]
\includegraphics[height=0.35\textheight]{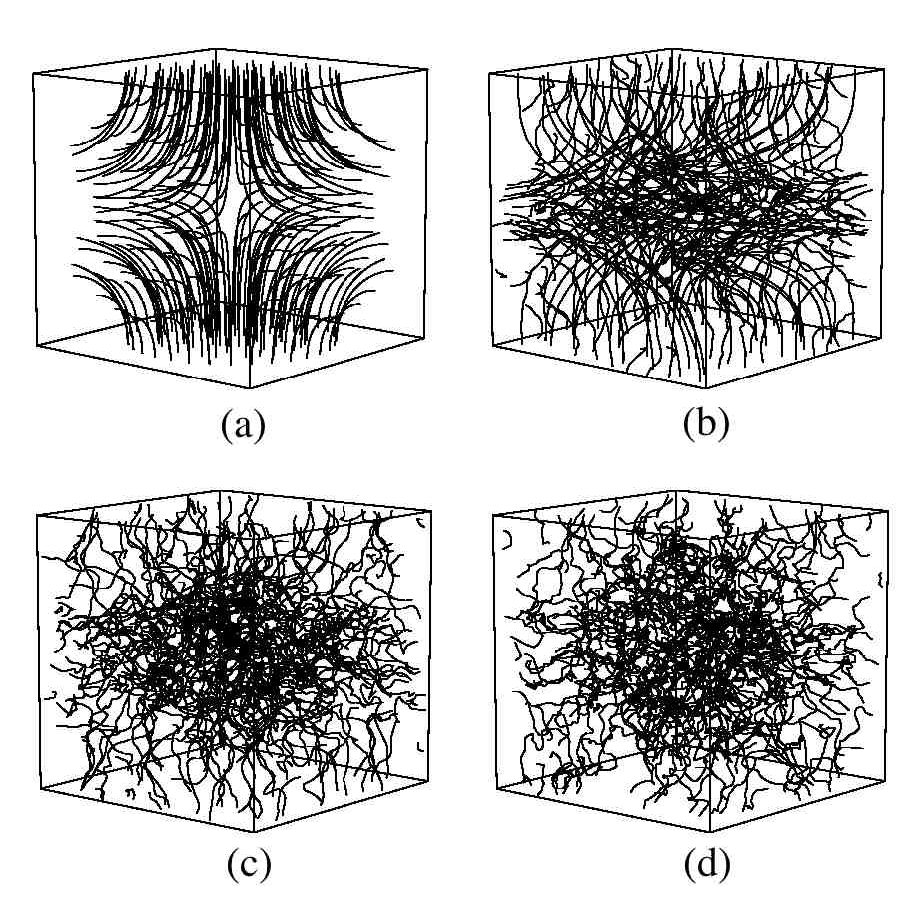}
\caption{Time development of the vortex tangle at $t=0$ sec (a), 30.0 sec (b),
50.0 sec (c) and 70.0 sec (d). The vortices are confined in a cube of size
$D=1$ cm and the calculation is made under the fully Biot-Savart law (Araki,
Tsubota and Nemirovskii, Phys. Rev. Lett. 89,145301-2, 2002, reproduced w
ith permission. Copyright(2002) by the American Physical Society).}
\end{figure}
Starting from the Taylor-Green vortex and following the vortex motion
without the mutual friction, we obtained a roughly homogeneous and isotropic
vortex tangle (Fig. 1) \cite{araki}. This is a decaying turbulence, being
dissipated by the cutoff of the smallest vortices whose size is comparable
to the space resolution $\Delta \xi=1.83 \times 10^{-2}$ cm. At first, the
energy spectrum has a large peak at the largest scale where the energy is
concentrated, but the spectrum changes as the vortices become homogeneous
and isotropic. The time dependence of the energy dissipation rate $\epsilon$
shows that $d\epsilon/dt$ becomes small after about 70 sec and artifacts of
the initial state disappear \cite{araki}. Similarly, the isotropic
parameters introduced by Schwarz \cite{Schwarz} indicate a nearly isotropic
vortex tangle becomes after 70 sec. Figure 2 shows that the energy spectrum
of the vortex tangle at 70 sec agrees quantitatively with the Kolmogorov
spectrum in the small $k$ region. The dissipative mechanism due to the
cutoff works only at the largest wave number $k \sim 2 \pi/\Delta \xi=343$
cm$^{-1}$. However the energy spectrum at small $k$ region is determined by
that dissipation rate.  By monitoring the development of the vortex size
distribution,  such decay of a tangle is found to be sustained by the
Richardson cascade process \cite{tsubota00}. These results support the
classical picture of the inertial range in superfluid turbulence at very low
temperatures.
\begin{figure}[btp]
\includegraphics[height=0.24\textheight]{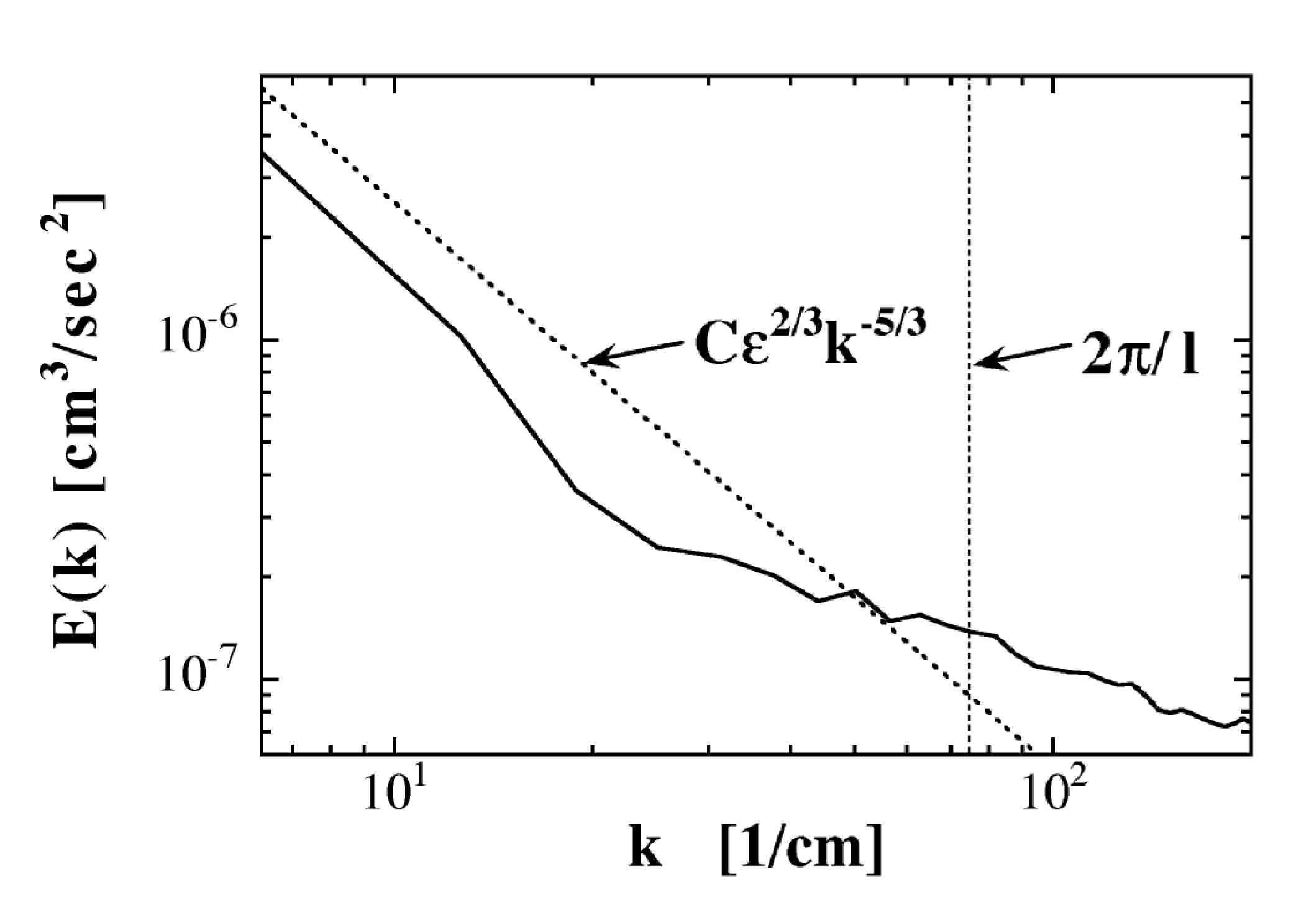}
\caption{Comparison of the energy spectrum (solid line) at $t=70$ sec with the
Kolmogorov law $E(k)=C\epsilon^{2/3}k^{-5/3}$ (dotted line) with $C=C_K=1$
and $\epsilon=1.287 \times 10^{-6}$ cm$^2$/sec$^3$ law (Araki, Tsubota and
Nemirovskii, Phys. Rev. Lett. 89, 145301-3, 2002, reproduced with
permission. Copyright(2002) by the American Physical Society). }
\end{figure}

The power of the spectrum changes from -5/3 to -1 at about $k \sim
2\pi/\ell$. The $k^{-1}$ spectrum comes from the contribution of the
velocity near each single vortex. Our simulation of this $k^{-1}$ regime
\cite{VTM} indicates that the energy cascade process is due to the Kelvin
wave cascade process. The simulations showed that when there is continuous
excitation of small $k$ Kelvin waves along a single vortex and if a sink
removes the energy at large wave numbers, then the nonlinear coupling
between different modes leads to a net flow of energy from small to large
wave numbers. This results in a simple steady spectrum of Kelvin waves that
is insensitive to the strength and frequency of the excited drive.

A specific picture of vortex tangles at very low temperatures has emerged
from these studies. In the low $k$ regime, the vortex tangle has the
inertial range in which the Kolmogorov spectrum is realized and the
Richardson cascade process describes the energy flow to larger wave numbers;
in this regime, superfluid turbulence mimics classical turbulence. In the
regime with $k \geq  2\pi/\ell$, the Kelvin wave cascade process becomes
relevant to the transfer of energy to the high wave numbers above which some
other mechanism dissipates the energy; this cascade process may be obscure
in classical turbulence because the wave numbers become comparable to the
vortex core sizes, but it is clearly seen in superfluid turbulence at very
low temperatures.

\subsection{DIFFUSION OF AN INHOMOGENEOUS VORTEX TANGLE}
At finite temperatures, vortex tangles decay due to mutual friction. We ask
here if vortex tangles decay at very low temperatures, where the normal
fluid component is so negligible that the mutual friction does not work, and
ask what mechanism removes energy from the tangle if it does decay.
Experimentally, these questions are not easy to answer because counterflow
and second sound techniques cannot work at such low temperatures.
Nevertheless, Davis {\it et al.} created vortices in superfluid $^4$He using
a vibrating grid, and observed their temperature-independent decay below $T
\sim 70$ mK by the trapping of negative ions at vortices \cite{Davis}. The
energy cascade process described in the previous subsection may be closely
related to this decay, although an understanding of the behavior of an
inhomogeneous vortex tangle is likely needed. Thus, we studied numerically
the diffusion of a localized tangle \cite{TAV}. We first prepared a
localized vortex tangle. Then we followed its decay and diffusion under the
fully Biot-Savart law in the absence of mutual friction.

The decay of a homogeneous vortex tangle is described by Vinen's equation
$dL/dt = -\chi_2 (\kappa/2\pi) L^2$ \cite{vinen3}. Here $L=(1/\Lambda) \int
d\xi$ is the vortex line density, where $\xi$ is the one-dimensional
coordinate along the vortex filaments and the integral is taken along all
vortices in the sample volume $\Lambda$. The parameter $\chi_2$ depends on
temperature, being about 0.3 at zero temperature \cite{tsubota00}. A simple
generalization of this equation for an inhomogeneous system may be obtained
by adding a diffusion term, so that we have the inhomogeneous Vinen's
equaion
\begin{equation}
\frac{L(\vr, t)}{dt} = -\chi_2 \frac{\kappa}{2\pi} L(\vr, t)^2 + D \nabla^2
L(\vr, t),
\label{IVE}
\end{equation}
where $D$ is a diffusion constant. A comparison between the numerical
solution of Eq. (\ref{IVE}) and the above numerical simulation shows that
$D=(0.1 \pm 0.05) \kappa$.  Hence the diffusion of a tangle is relatively
small; similar results were obtained by Barenghi and Samuels \cite{BS02}.
Even if a tangle is created by a vibrating grid, the diffusion is small and
the tangle can remain localized in the neighborhood of the grid as far as
there is no motion on a length scale larger than the line spacing. If motion
on a larger scale is generated, then the diffusion can be enhanced.

\subsection{ROTATING SUPERFLUID TURBULENCE}
Rotating superfluid turbulence is interesting because the rotation can cause
vortex ordering  and thus lead to various vortex phases.  This problem is
different from that of superfluid turbulence at zero temperature, but
closely related with the motivation becoming very important recently.  Most
configurations of quantized vortices which have been investigated in
superfluid helium can be grouped into two types: ordered vortex arrays and
disordered vortex tangles, and the former part of this article has been
devoted to the disordered vortex tangle. For example, ordered arrays of
vortices occur when superfluid helium is
rotated with angular velocity $\Omega$ exceeding a certain critical value
\cite{donnelly}. The resulting quantized vortices are aligned along the
rotation axis and form an array whose areal number density is
$2\Omega/\kappa$, in agreement with a rule discovered by Feynman. Thus, an
important question arose: what happens if vortices are created by {\it both}
rotation and counterflow along the rotational axis? Only one experiment
seems to address this issue\cite{SBD}. In the experiment, Swanson, Barenghi,
and Donnelly mounted a counterflow channel on a rotating cryostat, thus
being able to create vortices by independent combination of rotation and
counterflow. The vortex line density $L$ was determined from the
second-sound attenuation across the channel. They found two critical
velocities with $V_{c1} <V_{c2}$. At the lowest rotation velocities, when
the counterflow velocity $V_{\rm  ns}$   is less than $V_{c1}$, a vortex
array formed with a density in agreement with Feynman's rule. This critical
value of $V_{c1}$ was much less than that with no applied rotation. In
addition, the measured $V_{c1}$ was consistent with the onset of a vortex
wave instability discovered experimentally by Cheng {\it et al.}
\cite{Cheng}  and explained by Glaberson {\it et al.} \cite{Glaberson} as
being caused by helical Kelvin waves along the vortex cores that are
destabilized by the component of counterflow velocity along the vortices.
This instability is usually called the Donnelly-Glaberson (DG) instability.
The nature of the flow within this unstable regime and the cause of the
second critical velocity $V_{c2}$ is not clear.

To better understand these flow regimes, we simulated the original
experiments of Swanson {\it et al.} \cite{SBD}, in which the superfluid
flowed in a rotating cube, and found that the ordering of the vortex array
was sensitive to the rotation velocity. In a rotating cube, the equation of
motion of vortices is modified by two effects. The first effect is the force
acting upon the vortex due to the rotation. According to Helmholtz's
theorem, the generalized force acting upon the vortex is balanced by the
Magnus force as $\rho_{\rm s} \kappa (\vs' \times\dot{\vs}_0)=\delta
F'/\delta \vs $, where $\rho_{\rm s}$ is the superfluid density, $F'=F-\vOme
\cdot \vM $ is the free energy of a system in the frame rotating with the
angular velocity $\vOme$, and $\vM$ is the angular momentum. Taking the
vector product with $\vs'$, we obtained the velocity $\dot{\vs}_0$. The
first term $F$ due to the kinetic energy of vortices gives the Biot -Savart
law of Eq. (\ref{sdot}). The second term $\vOme \cdot \vM $ leads to the
velocity $\dot{\vs}_{\rm rot}$ of the vortex caused by the rotation. The
second effect is the superflow induced by the rotating square cross-section.
For a perfect fluid, we can use the analytical solution of the velocity
$\vv_{\rm cub}$ inside a cube of size $D$ rotating with angular velocity
$\vOme=\Omega \hat{\vz}$ \cite{Thomson}. By including both effects, we
obtained the velocity $\dot{\vs}_0$ in a rotating cube as the sum of
$\vv_{\rm cub}$ and $\dot{\vs}_{\rm rot}$, the right-hand side of Eq.
(\ref{sdot}).  In the numerical simulation, the sample was a cube of size
$D=1.0$ cm. We used periodic boundary conditions along the rotating axis and
rigid boundary conditions at the side-walls. Also, the counterflow was
applied along the $z$-axis, the normal fluid was assumed to be at rest in
the rotating frame, and, to make comparison with the experiment \cite{SBD},
we did the calculation for a temperature $T=1.6$ K. Hence, the calculation
included mutual friction. Two quantities were introduced to characterize the
vortex tangle. One is the vortex line density $L$, and the other is the
polarization of the tangle which is defined as $\langle s'_z
\rangle=(1/\Lambda L) \int d\xi {\vs}' \cdot \hat{\mbox{\boldmath $z$}}$.
Thus $\langle s'_z \rangle$ is unity for a perfect vortex array and zero for
a randomly oriented tangle.

The numerical simulation began from a rotating vortex array with small
random perturbations added to the vortices, then a
counterflow velocity was applied. Figure 3 shows how 33 vortices, initially
parallel to each other in a stable configuration at $\Omega=4.98 \times
10^{-2}$ rad/sec, develop into a vortex tangle under  $V_{\rm ns}=0.08$
cm/sec. The parallel vortices become unstable and Kelvin waves grow by the
DG instability. When the amplitude of the Kelvin waves becomes comparable to
the average vortex separation, vortex reconnections take place and more
vortex loops are created. These loops disturb the initial vortex array,
leading to an apparently random vortex tangle.
\begin{figure}[btp]
\includegraphics[height=0.37\textheight]{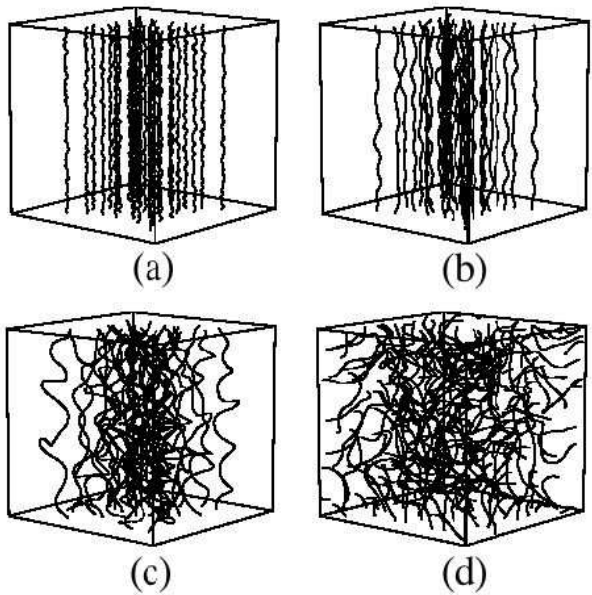}
\caption{Development from a vortex array to a polarized tangle at $T=1.6$ K,
$\Omega=4.98 \times 10^{-2}$ rad/sec and $V_{ns}=0.08$ cm/sec.  Computed
vortex tangle at the following times: (a): t=0 sec; (b): 12sec; (c): 28 sec;
(d): 160 sec (Tsubota, Araki and Barenghi, Phys. Rev. Lett. 90, 145301-3,
2003, reproduced with permission. Copyright(2003) by the American Physical
Society). }
\end{figure}

The results show that rotation tends to order the vortex tangle over that
which would occur without rotation \cite{Schwarz,tsubota00}.  The values of
$L$ and $\langle s'_z \rangle$ show that the vortex tangle evolves into a
statistically saturated steady state. A vortex tangle created with only
counterflow is sustained by the competition between the driving counterflow
and the mutual friction \cite{Schwarz}, whereas in our case the steady state
is achieved by balancing the effects of rotation, counterflow, and mutual
friction. Figure 4 shows the dependence of $L$ and $\langle s'_z \rangle$ on
$V^2_{\rm ns}$ and $\Omega$ in each saturated state. It is evident that the
polarization decreases with increasing counterflow velocity and increases
with increasing rotation. This shows the competition between ordering caused
by rotation and disordering caused by counterflow.
\begin{figure}[btp]
\includegraphics[height=0.17\textheight]{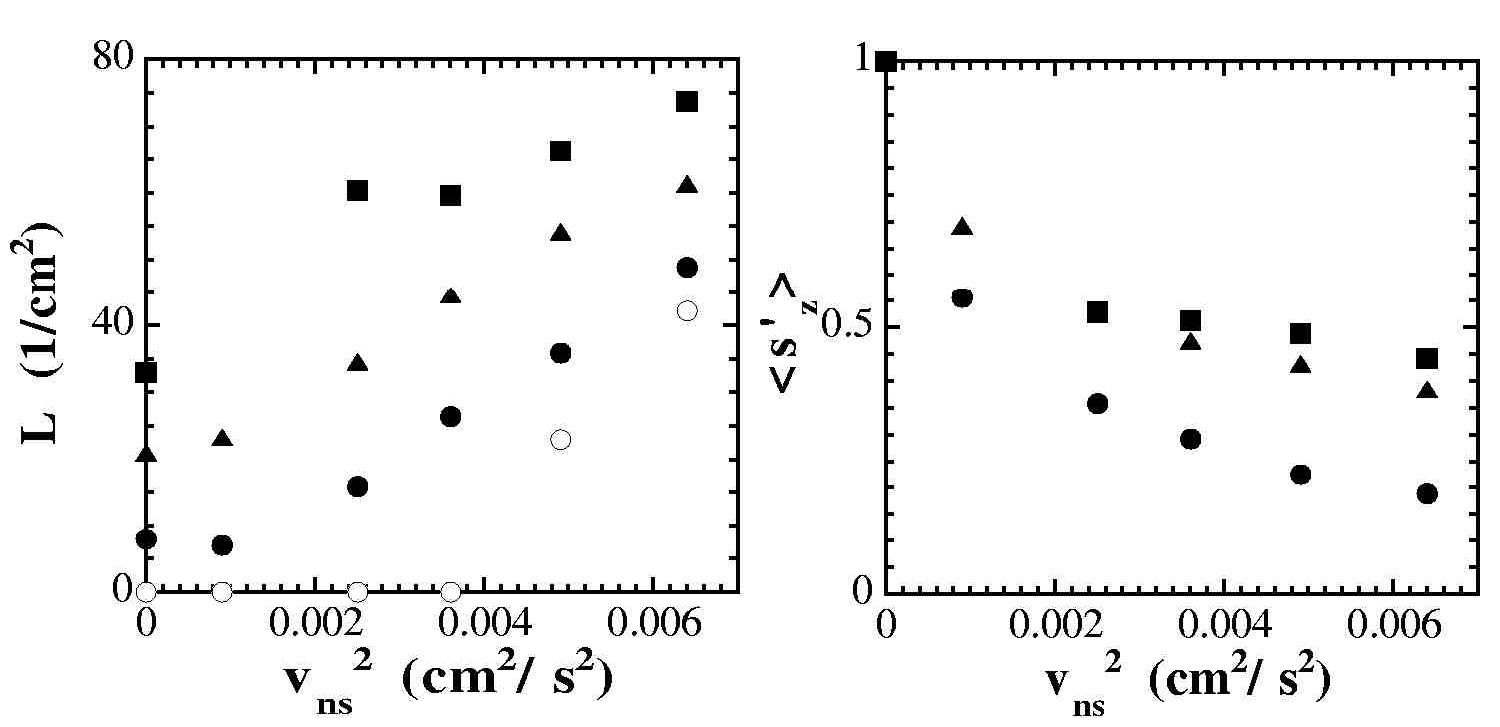}
\caption{Dependence of $L$ and $\langle s'_z \rangle$ on $v^2_{ns}$ in the
saturated state for $\Omega=0$ (white square), $\Omega=9.97 \times 10^{-3}$
rad/sec (black square), $\Omega=2.99 \times 10^{-2}$ rad/sec (circle) and
$\Omega=4.98\times 10^{-2}$ rad/sec (triangle). The right figure lacks the
data for $\Omega=0$ (Tsubota, Araki and Barenghi, Phys. Rev. Lett. 90,
145301-3, 2003, reproduced with permission. Copyright(2003) by the American
Physical Society). }
\end{figure}

Recently Finne {\it et al.}  used NMR to study turbulence of the B phase of
superfluid $^3$He and observed a sharp velocity-independent transition at a
critical temperature between two regimes \cite{finne}. Regular behavior
occurred at high temperature and turbulence occurred at low temperatures.
They also found that the experimental results were consistent with numerical
vortex dynamics simulations which was done by the method described here.
This turbulence comes from intrinsic instability of vortices depending on
temperature,  being different from the DG instability.

\section{DYNAMICS IN ROTATING BOSE-EINSTEIN CONDENSATE}\label{roteBEC}
\subsection{QUANTIZED VORTICES IN ALKALI-ATOMIC BOSE-EINSTEIN CONDENSATE}
The recent dramatic achievement of Bose-Einstein condensation in trapped
alkali-atomic gases at ultra-low temperatures has stimulated intense
experimental and theoretical activity \cite{ps}. Such atomic-gas
Bose-Einstein condensates (BECs) differ fundamentally from liquid helium
condensates in several ways. First, the condensates of alkali-atomic gases
are dilute, having the mean particle density $n$ with $n|a|^3\ll 1$,  so
that the interatomic interaction can be accurately parametrized in terms of
a scattering length $a$. As a result, at low temperatures, the GP equation
(\ref{gpeq}) gives an extremely precise description of the atomic condensate
and their dynamics. This situation differs from superfluid $^4$He, where the
relatively high density and strong repulsive interactions complicate greatly
the analytical treatments. Furthermore, because of the relatively strong
interactions, the condensate fraction in bulk superfluid $^4$He is only
about 10\%\ of the total particles, even at zero temperature. In contrast,
almost all atoms participate in the condensate in an atomic-gas BEC. Second,
a BEC in bulk helium has uniform density, whereas an atomic-gas BEC has
nonuniform density due to the confinment by a trapping potential. In this
section, we review the dynamics of quantized vortices in an atomic-gas BEC
system.

 From the above properties, the physics of quantized vortices has novel
features. First, since the density is dilute, the relatively large coherence
length $\xi$ makes it possible to directly visualize the quantized vortices
by using optical techniques. Second, because the order of the coherence
length $\xi$ is close to the size of the condensate, the vortex dynamics is
closely connected with the collective motion of the condensate density
itself. As described later, vortex nucleation of this system is related to
an instability in the surface excitations of the condensate, which differs
from that in superfluid helium system where the compressibility of the
density is probably negligible. Finally, alkali atoms have internal degrees
of freedom attributed to the hyperfine spin, so that their BECs can have
multi-component order parameters if the spin degrees are available.
Multicomponent BECs provide us with new possibilities to study
unconventional vortex states that have been studied extensively in other
fields of physics such as superfluid $^{3}$He, anisotropic superconductors,
and cosmology.

Superfluids can support dissipationless flow, which is closely related to
the existence of quantized vortices. Hence, the importance of studying a
quantized vortex was recognized immediately after the first atomic-gas BEC
was realized, in order to know whether or not the system becomes superfluid.
The first experimental detection of a vortex involved a $^{87}$Rb BEC
containing two different internal hyperfine components \cite{Matthew}. The
generation of a vortex was achieved by a phase engineering technique, in
which interconversion  between two components was controlled spatially and
temporally by an external coupling field \cite{Williams}. They then turned
off the coupling, thus creating a vortex state consisting of one circulating
component surrounding a nonrotating core of the other component. The ENS
group observed the formation of single and multiple vortices in a
single-component, $^{87}$Rb elongated cigar-shape condensate \cite{Madison}.
The condensate was trapped in a static axisymmetric magnetic trap and a
nonaxisymmetric attractive dipole potential created by a stirring laser
beam. This combined potential produces a cigar-shaped harmonic trap with a
slightly anisotropic transverse profile. By rotating the transverse
anisotropy at a frequency $\Omega$, they could observe the formation of a
vortex above a certain critical value of $\Omega$. A lattice consisting of
more vortices appeared when $\Omega$ was increased further. In contrast to
the indirect visualization methods used in superfluid helium systems, the
quantized vortices were directly visualized as holes in the transverse
density profile of the time-of-flight images. Later, groups at MIT
\cite{Abo}, JILA \cite{Haljan}, and Oxford \cite{Hodby} observed vortex
lattices; the first two groups observed lattices that consisted of 100 or
more vortices.

There are numerous theoretical studies of quantized vortices in atomic-gas
BECs (see Ref. \cite{Fetter}). In this area, our group focused on the
dynamical properties of quantized vortices.  Based on the numerical
simulation of the time-dependent GP equation, we revealed the marvelous
nonlinear dynamics that a rotating BEC undergoes
\cite{TsubotaBEC,Kasaken2,Kasaken}. In Sec. \ref{latdy}, we discuss the
dynamics of vortex lattice formation in a rotating BEC
\cite{TsubotaBEC,Kasaken},  motivated by the experimental observation of
Ref. \cite{Madison2}. This work clarified for the first time the nonlinear
dynamics of vortex lattice formation, in good agreement with the
experimental results. In Sec. \ref{gianvor}, we study the dynamics of a fast
rotating condensate \cite{Kasaken2}. By combining an additional quartic
potential with a harmonic potential, we find the new vortex states involving
``a giant vortex'', in which many vortices are absorbed in a single density
hole.

\subsection{VORTEX LATTICE FORMATION IN A ROTATING BOSE-EINSTEIN CONDENSATE}
\label{latdy}
  Madison {\it et al}. observed directly nonlinear dynamical phenomena such
as vortex nucleation and lattice formation in a rotating condensate
\cite{Madison2}. By suddenly turning on the rotation of the potential, the
initially axisymmetric condensate made a collective quadrupole oscillation
in which the condensate deformed elliptically. This oscillation continued
for a few hundred milliseconds with a gradually decreasing amplitude. Then,
the axial symmetry of the condensate suddenly reappeared and concurrently
the vortices entered the condensate from its surface, eventually settling
into a lattice configuration in the bulk.
Such a direct observation of vortex nucleation and vortex lattice formation
has never been done in superfluid helium. This observation motivated us to
study theoretically the detailed dynamics of vortex lattice formation.

One motivation for studying a rotating, atomic-gas BEC is to determine
whether the dynamical instability \cite{Sinha}, the Landau instability
\cite{Dalfovo}, or both are responsible for vortex nucleation and lattice
formation. The dynamical instability originates from the imaginary frequency
of the excitation mode, giving rise to an exponential growth of the unstable
mode even in the energy-conserving dynamics. The Landau instability occurs
when the excitation spectrum has negative eigenvalues in the rotating frame
and the system is subject to some energy dissipative mechanism. These two
instabilities usually occur in different parameter regimes. Thus, one may
ask which instability is important for actual vortex nucleation. In the
experiments \cite{Madison,Madison2,Abo}, the vortices nucleated most
frequently when the rotation frequency $\Omega$ was near
$0.7\omega_{\perp}$, where $\omega_{\perp}$ is the transverse trapping
frequency. The rotating potential in these experiments excites mainly the
surface mode with the quantum number of angular momentum $l=2$ (quadrupole
mode). When the interaction energy of the condensate is much larger than the
energy of a harmonic potential, the dispersion relation for the surface mode
is given by $\omega_{l}=\sqrt{l}\omega_{\perp}$ \cite{Stringari}. In the
rotating frame with frequency $\Omega$, the surface-mode frequency is
shifted by $-l\Omega$. Hence, it is expected that the quadrupole mode $l=2$
is resonantly excited at $\Omega/\omega_{\perp}=\sqrt{2}/2 \simeq 0.707$.
Sinha and Castin showed that the rotating condensate without vorticity was
dynamically unstable near the quadrupole resonance and proposed that this
instability triggered vortex nucleation \cite{Sinha}. However, our later
analysis reveals that both instabilities are necessary to explain the actual
vortex formation.

We resolved these problems by numerically solving the GP equation which is
an extension of Eq. (\ref{gpeq}) \cite{TsubotaBEC,Kasaken}
 \begin{eqnarray}
 (i - \gamma)\hbar \frac{\partial \psi}{\partial t} = \Bigl[
-\frac{\hbar^2}{2m}\nabla^2+ V -\mu + g |\psi|^2  -\Omega L_z \Bigr] \psi.
\label{geneGPE}
\end{eqnarray}
Here, the centrifugal term $-\Omega L_z=i \hbar \Omega (x\partial_y -
y\partial_x)$ appears in a system rotating about the $z$-axis at a frequency
$\Omega$. In this work, we assume translational symmetry along the $z$-axis,
thus making the problem two-dimensional. Then the normalization of the wave
function $\psi$ is taken as $\int d {\bf r} |\psi|^{2}=n_{\rm 2D}$ with the
particle number per unit length along the $z$-axis. The trapping potential
has the form
\begin{equation}
V ({\bf r})= \frac{1}{2} m \omega_{\perp}^{2} \{
(1+\epsilon_x)x^2+(1+\epsilon_y)y^2 \},
\label{trappotential}
\end{equation}
with the small anisotropy parameters $\epsilon_{x}$ and  $\epsilon_{y}$
($\epsilon_{x} \neq \epsilon_{y}$) ; this form describes approximately the
rotating potential used in the ENS experiments \cite{Madison,Madison2}. By
introducing the scales characterizing the trapping potential
$a_{h}=\sqrt{\hbar/2 m \omega_{\perp}}$ for length and $\omega_{\perp}^{-1}$
for time, and replacing $\psi \rightarrow \sqrt{n_{\rm 2D}}\psi/a_{h}$, the
coupling coefficient in the nonlinear term of Eq. (\ref{geneGPE}) becomes
$C=8 \pi a n_{\rm 2D}$. The condition of Ref. \cite{Madison2}, corresponding
to $a=5.77$nm, $N=3 \times 10^{5}$, $\omega_{z}=11.8 \times 2 \pi$ and
$\lambda=\omega_{\perp}/\omega_{z}=9.2$, gives $C \simeq 500$. The term with
$\gamma$ introduces dissipation, which is treated phenomenologically in the
GP equation. This form of the dissipative equation follows the work of Choi
{\it et al.} \cite{Choi};  they studied the damping of the collective
oscillation of a BEC and determined the value of $\gamma$ to be 0.03 by
fitting their theoretical results with the MIT experiments \cite{Mewes}.
Hence, we use $\gamma =0.03$ here. Because this dissipative term is much
smaller than other terms in the GP equation, a small variation of $\gamma$
would not change the dynamics qualitatively but can only modify the
relaxation time scale. This form can be derived from the formulation for a
finite-temperature BEC developed by Zaremba {\it et al.} \cite{Zaremba}
under some approximations \cite{Kasaken}. 
Gardiner {\it et al.} obtained the similar but different equation by another approach 
\cite{Gardiner}, applying it to the simulation of vortex lattice formation \cite{Penkwitt}. 
Because the time development of Eq. (\ref{geneGPE}) does not conserve the norm 
of the wave function, we had to adjust the chemical potential $\mu$ at each time step 
to ensure normalization.

\begin{figure}[btp]
\includegraphics[height=0.55\textheight]{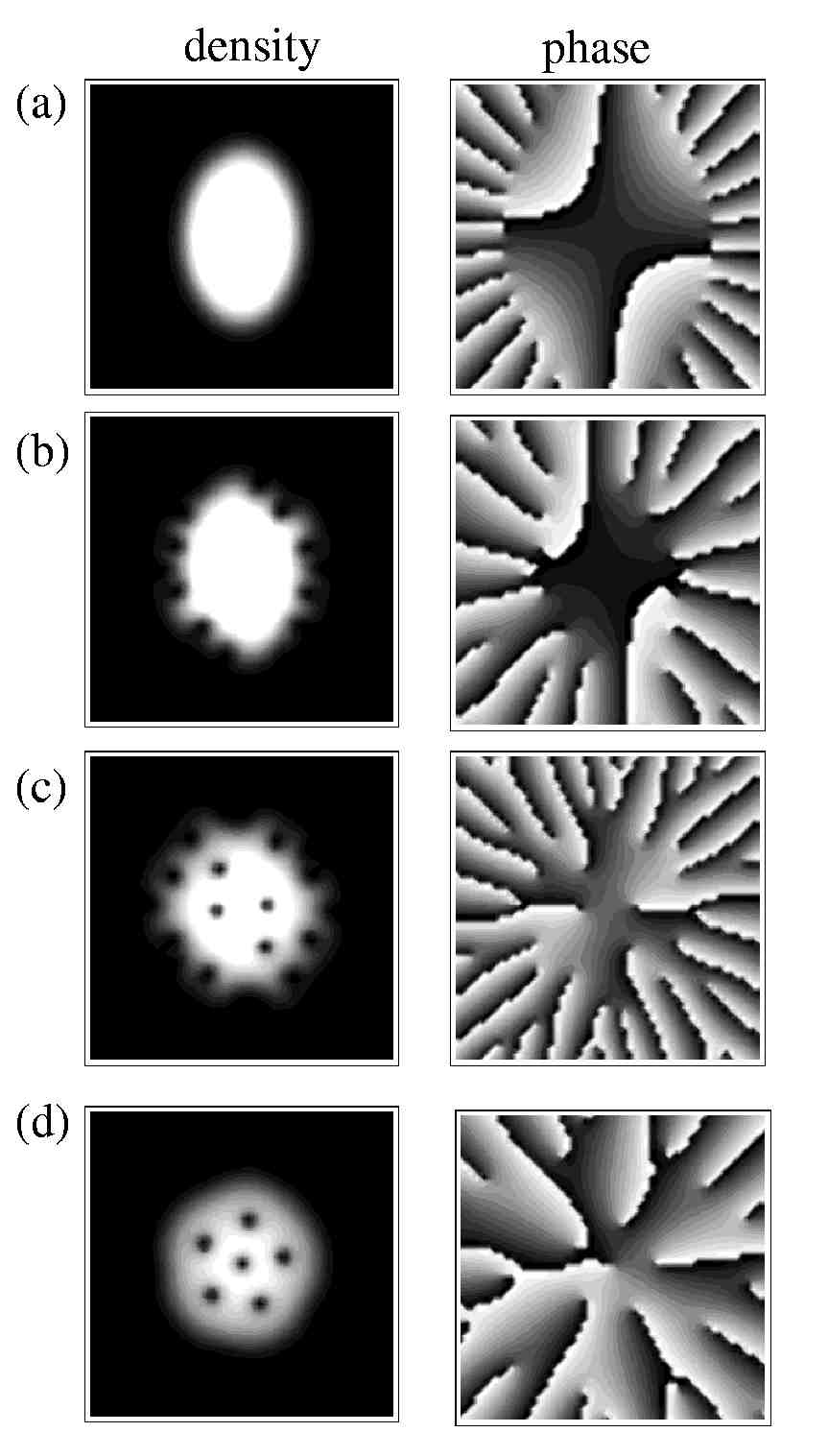}
\caption{Time development of the density (left column) and phase (right column)
profile after the trapping potential begins to rotate suddenly with
$\Omega=0.7\omega_{\perp}$ for (a) 67 msec, (b) 360 msec, (c) 425 msec and
(d) 735 msec. The value of the phase varies continuously from 0 (black) to
2$\pi$ (white). There appear some lines where the phase changes
discontinuously from black to white, corresponding to the branch cuts
between the phase 0 and $2\pi$. The apexes of these lines around which the
value of the phase rotates continuously from 0 to $2\pi$ represent the
quantized vortices. }
\end{figure}
The simulations started from the stationary solution with uniform phase of
Eq. (\ref{geneGPE}) for a nonrotating trap ($\epsilon_{x}=\epsilon_{y}=0$).
The method for turning on a rotating drive followed the experimental
procedure in Ref. \cite{Madison2}. The rotation with frequency $\Omega$
started at $t=0$, and the trap anisotoropy $\epsilon = \{ (1+\epsilon_{x}) -
(1+\epsilon_{y}) \}/ \{ (1+\epsilon_{x}) + (1+\epsilon_{y}) \}$ was
increased rapidly from zero to its final value 0.025 in 20 msec,  keeping
$\epsilon_{y}$ zero. First, we discuss the dynamics of the condensate with
$C=500$ for $\Omega/\omega_{\perp}=0.7$. When the dissipation is free
($\gamma=0$), the condensate undergoes a quadrupole oscillation, but it
shows only the simple periodic oscillation in a sense of the
Fermi-Pasta-Ulam recurrence \cite{Fermi} [see Fig. 6]. When the dissipation
is present, this behavior is dramatically changed. The left column of Fig. 5
shows the time development of the condensate density $|\psi(x,y,t)|^2$.
Initially, the condensate also undergoes a quadrupole oscillation, but the
oscillation is damped because of the dissipation. After a few hundred
milliseconds, the boundary surface of the condensate becomes unstable,
generating surface ripples that propagate along the surface. The excitations
are likely to occur on the surface whose curvature is low, i.e. parallel to
the longer axis of the ellipse as shown in Fig. 5(b). Then the waves on the
surface develop into the vortex cores, which enter the condensate. As the
vortices penetrate inside the condensate, the axisymmetry of the condensate
shape is recovered. As is known in the study of rotating superfluid helium
\cite{Tsubota95}, the rotating drive pulls vortices toward the rotation
axis, whereas repulsive interactions between vortices tends to push them
apart; therefore, their competition yields a vortex lattice whose vortex
density  depends on the rotation frequency. In this simulation, six vortices
form a vortex lattice. Figure 6 shows the time development of the
deformation parameter defined as
\begin{equation}
\alpha(t)=-\Omega \frac{\langle x^{2} \rangle -\langle y^{2}
\rangle}{\langle x^{2} \rangle + \langle y^{2} \rangle},
\label{deforma}
\end{equation}
where $\langle A \rangle$ means $\int dxdy \psi^{\ast} A \psi$. Madison {\it
et al.} observed that, after a rotation of $\Omega = 0.7\omega_{\perp}$
started suddenly,  $\alpha$ oscillated during a few hundred milliseconds and
then reduced abruptly to a value below 0.1 when vortices entered the
condensate from its surface \cite{Madison2}. The evolution of $\alpha$
(bold-solid curve in Fig. 6) closely resembles Fig. 3 of Ref.
\cite{Madison2},  and our result is thus consistent with the experimental
one.
\begin{figure}[btp]
\includegraphics[height=0.22\textheight]{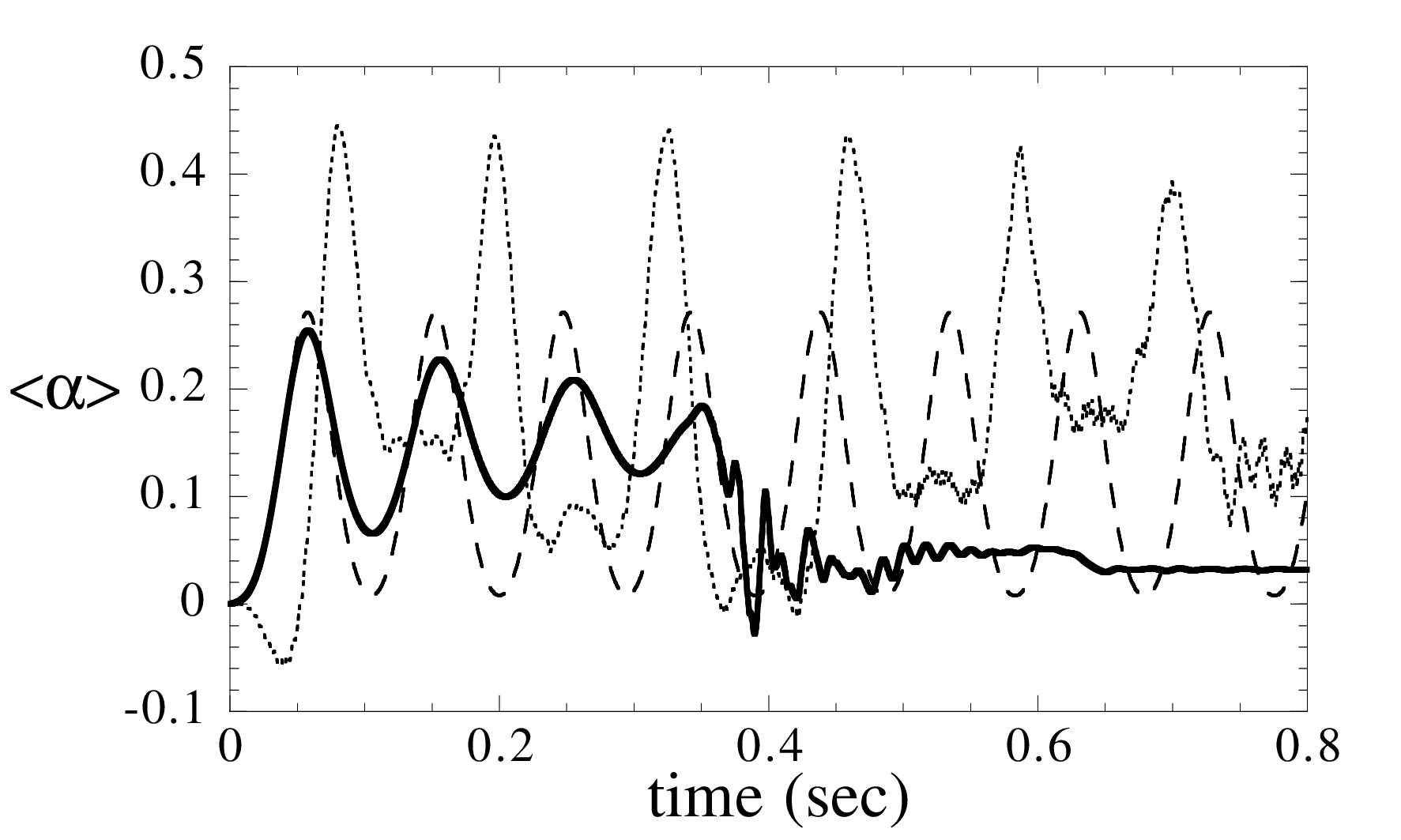}
\caption{Time evolution of the deformation parameter $\alpha$ for
$\Omega/\omega_{\perp}=0.70$ with dissipation (bold-solid curve), for
$\Omega/\omega_{\perp}=0.70$ without dissipation (dashed-curve) and for
$\Omega/\omega_{\perp}=0.75$ without dissipation (dotted-curve).  }
\end{figure}

The corresponding time development of the phase of $\psi(x,y,t)$ is shown in
the right column of Fig. 5. As soon as the rotation starts, the phase field
inside the condensate takes the form of quadrupolar flow
$\theta(x,y)=\alpha(t) x y + {\rm const}$, and just outside the Thomas-Fermi
boundary appear many proto-vortices; for example, Fig. 5(a) shows about 20
proto-vortices. When the proto-vortices are on the outskirts of the
condensate where the amplitude $|\psi({\bf r},t)|$ is almost negligible,
they neither contribute to the energy nor the angular momentum of the
system. Because they are invisible in the corresponding density profile of F
ig. 5, they may be called ``ghost vortices". The ghost vortices move toward
the rotation axis, but their invasion into the condensate is prevented at
the boundary surface within which the amplitude grows up. However, as the
surface ripples are generated, the ghost vortices begin to penetrate the
condensate. At this point, only some defects enter the bulk condensate
because their further invasion costs energy. For example, Fig. 5(d) shows
that six vortices enter the condensate and form a lattice, while the
remaining vortices are repelled and escape to the outside of the
Thomas-Fermi boundary.

The penetration of the ghost vortices into the condensate was accomplished
with the help of the surface ripples, induced by the instabilities in the
non-vortex state. One possible mechanism for the surface ripple excitation
is the dynamical instability, which occurs without any dissipation. Sinha
and Castin did a linear stability analysis of a rotating condensate, finding
the growth of the fluctuation in the resonant range
$0.72<\Omega/\omega_{\perp}<0.78$ \cite{Sinha}. To confirm whether this
instability leads to the vortex lattice formation, we calculated the
dynamics for $\Omega/\omega_{\perp}=0.75$, and show the evolution of
$\alpha$ in Fig. 6. We can see that the perfect recurrence of the
oscillation is broken. This is due to the dynamical instability, which
causes the irreversible transfer of the quadrupole mode into the
higher-energy excitation mode via the nonlinear mode coupling. However,
although the surface ripples were generated, the ghost vortices never
penetrated the inside of the condensate. Therefore, the dynamical
instability alone cannot explain the vortex lattice formation. Actually, in
the experiments, the vortices occurred at even lower off-resonant
frequencies\cite{Madison,Abo,Hodby,Madison2}, a finding that cannot be
understood using the dynamical instability. The other instability is the
Landau instability, which is applicable to the case with dissipation. The
critical frequency of this instability can be expressed by the Landau
criterion $\Omega_{c}={\rm min} (\omega_l/l )$ for a rotating BEC
\cite{Dalfovo}. The angular momentum number $l_{c}$ that yields $\Omega_{c}$
takes a value larger than 4 with the parameter used in experiments
\cite{Dalfovo}. For $C=500$, we obtain $l_{c}$=8 and
$\Omega_{c}=0.5\omega_{\perp}$ from $\omega_{l}$ by numerically solving the
Bogoliubov-De Genne equation. Thus, our simulation confirmed that this
instability actually leads to the generation of vortices, where the surface
ripples always evolved into the vortex cores and penetrated into the
condensate.

In the range $0.72<\Omega/\omega_{\perp} <0.78$ where the dynamical
instability is effective, the surface ripples are excited, but dissipation
is needed for the vortices to penetrate into the condensate. Here the
dissipation may originate in the thermal component, which should be almost
negligible in the atomic gas experiments at very low temperatures. However,
in the dynamical process of a condensate, there is a possibility that the
thermal component will be produced under a strong perturbation. Hence the
experimental results of Ref. \cite{Madison,Madison2,Abo} may be explained as
follows. Consider a situation in which there is almost no dissipation at
very low temperatures, namely, a case in which the Landau instability cannot
occur. In a quadrupole resonance, however, the dynamical instability causes
stochasticity in condensate oscillations, leading to the creation of the
thermal component. Indeed, Hodby {\it et al.} has reported that the
temperature of the system increases from $0.5T_{c}$ to $0.8T_{c}$ when the
vortices form \cite{Hodby}. The time spent during this process is determined
by the growth time of the dynamical instability, which is expected to be
about 100 msec \cite{Sinha}. Thus, this oscillation-created thermal
component makes the dissipation effective, so the vortices can penetrate
into the condensate via the Landau instability. To test this hypothesis we
need to use analysis beyond the GP equation and thus we leave this issue for
future study. In the other range of $\Omega$, a condensate makes only a
stable quadrupole oscillation without dissipation. However, our results show
that vortices may be generated at frequencies above $\Omega_{c}$ whenever
finite dissipation occurs. Therefore, if one does an experiment in which the
temperature is high enough to have effective dissipation, one should be able
to observe the critical frequency given by the Landau criterion.

\subsection{GIANT VORTEX FORMATION IN A FAST ROTATING BOSE-EINSTEIN
CONDENSATE}\label{gianvor}
In this subsection, we consider the dynamics of a rotating BEC trapped in a
harmonic-plus-quartic potential \cite{Kasaken2}. The greater the rotation
frequency, the greater the number of vortices in the lattice. In this study,
we want to find out what happens when the size of the vortex cores becomes
comparable with the vortex separation. In such a situation, the vortex
lattice is predicted to melt via quantum fluctuation, leading to a quantum
Hall-like state \cite{Ho,Cooper,Sinova}. For a rotating condensate with
frequency $\Omega$ in a harmonic trapping potential
$(1/2)m\omega_{\perp}^{2}r^{2}$, the centrifugal potential cancels the
confinement, thus preventing a BEC from rotating at $\Omega$ beyond
$\omega_{\perp}$. Unfortunately, some authors have concluded that the study
of a dense vortex lattice encounters significant difficulties arising from
such a softening of the effective trap potential \cite{Fetter2,Fischer}.
This difficulty can be avoided by introducing an additional quartic
potential, so that the effective potential for the rotating condensate
becomes
\begin{equation}
V(r) = \frac{1}{2} m (\omega_{\perp}^{2}-\Omega^{2}) r^{2}  + \frac{1}{4} k
r^{4}.
\label{Mexic}
\end{equation}
The quartic part of $V$ allows us to increase $\Omega$ above
$\omega_{\perp}$, then the potential has a ``Mexican hat" structure. Hence,
we can expect the dynamics to be characteristic of the
quadratic-plus-quartic potential for $\Omega>\omega_{\perp}$. As in the last
subsection, we also consider a two-dimensional system subject to rotation
${\bf \Omega}=\Omega {\bf \hat{z}}$ by assuming translational symmetry along
the $z$ axis. In the numerical simulation, the rotating drive is suddenly
turned on by introducing a small anisotropy of the harmonic trap as $(
\epsilon_{x} x^{2} + \epsilon_{y} y^{2})/4$. Finally, the small trap
anisotropy $(\epsilon_x, \epsilon_y)$ is turned off adiabatically to obtain
the axisymmetric final steady state.

Numerical simulations were done for several values of $\Omega$ at fixed
$C=250$ and $k=1$. Figure 7 shows the time development of the density
profiles for $\Omega/\omega_{\perp}=2.5$ and $\Omega/\omega_{\perp}=3.2$. As
before, the ripples are excited on the condensate surface, and some ripples
develop into vortex cores and penetrate into the condensate. For
$\Omega/\omega_{\perp}=2.5$, however, some penetrating vortices move toward
the center, merging together to make a density hole at the center, around
which some other vortices form a circular array. The phase profiles of the
final steady state are shown in Fig. 7(d). Although some phase defects come
very close to each other in the central hole, they never overlap.
\begin{figure}[btp]
\includegraphics[height=0.55\textheight]{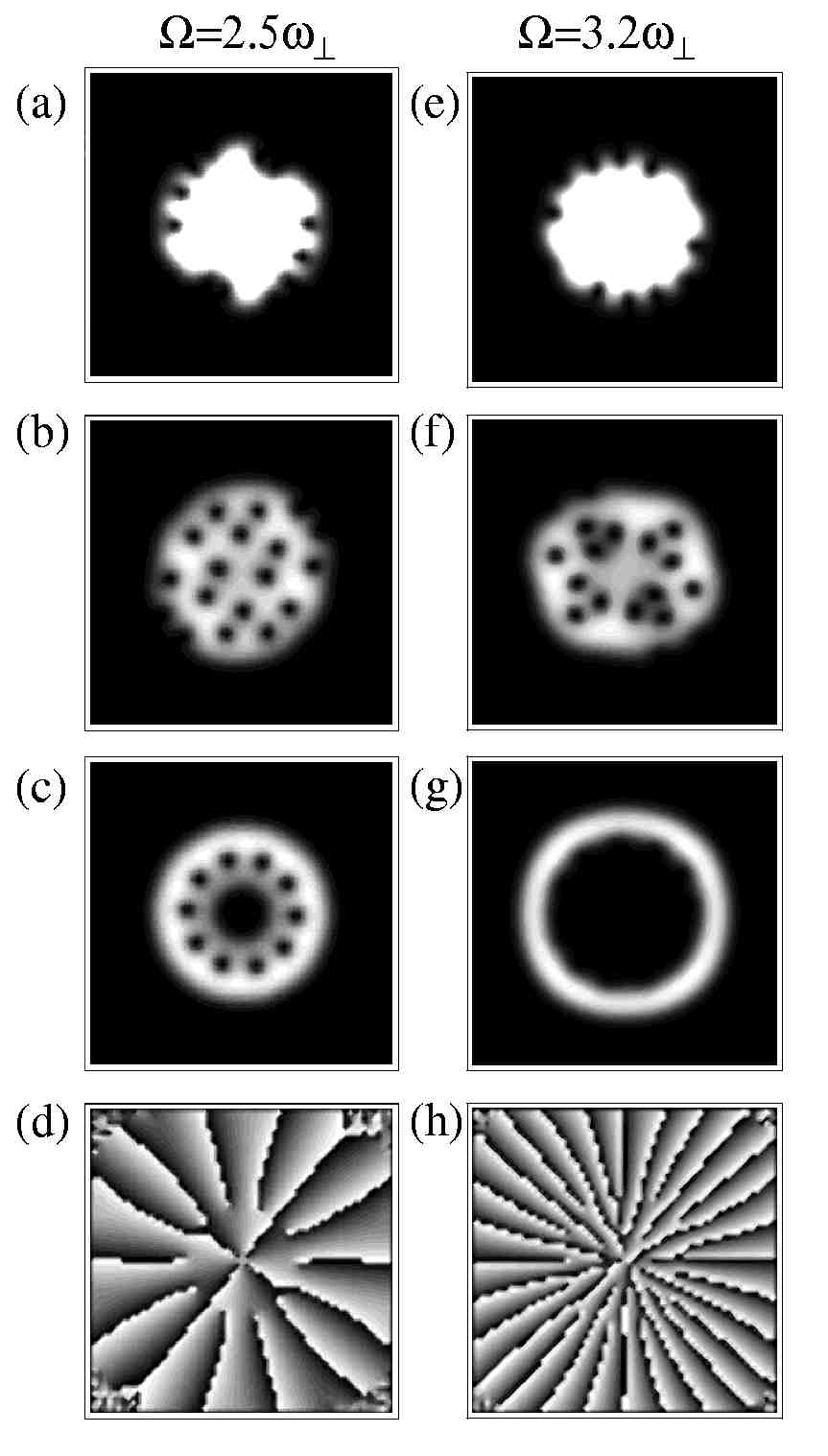}
\caption{Time development of the density profile for $\Omega=2.5\omega_{\perp}$
(left column) and $ \Omega=3.2\omega_{\perp} $(right column) after the
trapping potential begins to rotate suddenly. The time is (a) 30 msec, (b)
33 msec, (c) 300 msec for the left column and (e) 20 msec, (f) 23 msec, (g)
200 msec for the right column. The figure (d) and (h) shows the phase
profile corresponding to Fig. (c) and (g), respectively.}
\end{figure}

As $\Omega$ is further increased, all vortices generated from the condensate
surface are absorbed into the central density hole as shown in the right
column of Fig. 7, where the central hole is composed of 24 singly-quantized
vortices packed together. The packing is possible for high rotation
frequencies ($\Omega>\omega_{\perp}$) because the centrifugal force
decreases the central condensate density, and thus packing the vortices
together costs less energy. The minimum of Eq. (\ref{Mexic}) determines the
radius of the ring condensate in Fig. 7(g) as
$R=\sqrt{m(\Omega^{2}-\omega^2_{\perp} )/k}$. Here we call a set of vortices
such as that in Fig. 7(g) `` a giant vortex" to indicate that a number of
phase defects are contained in a single hole. Figures 7(h) also shows that
the phase singularities do not completely overlap.

\begin{figure}[btp]
\includegraphics[height=0.23\textheight]{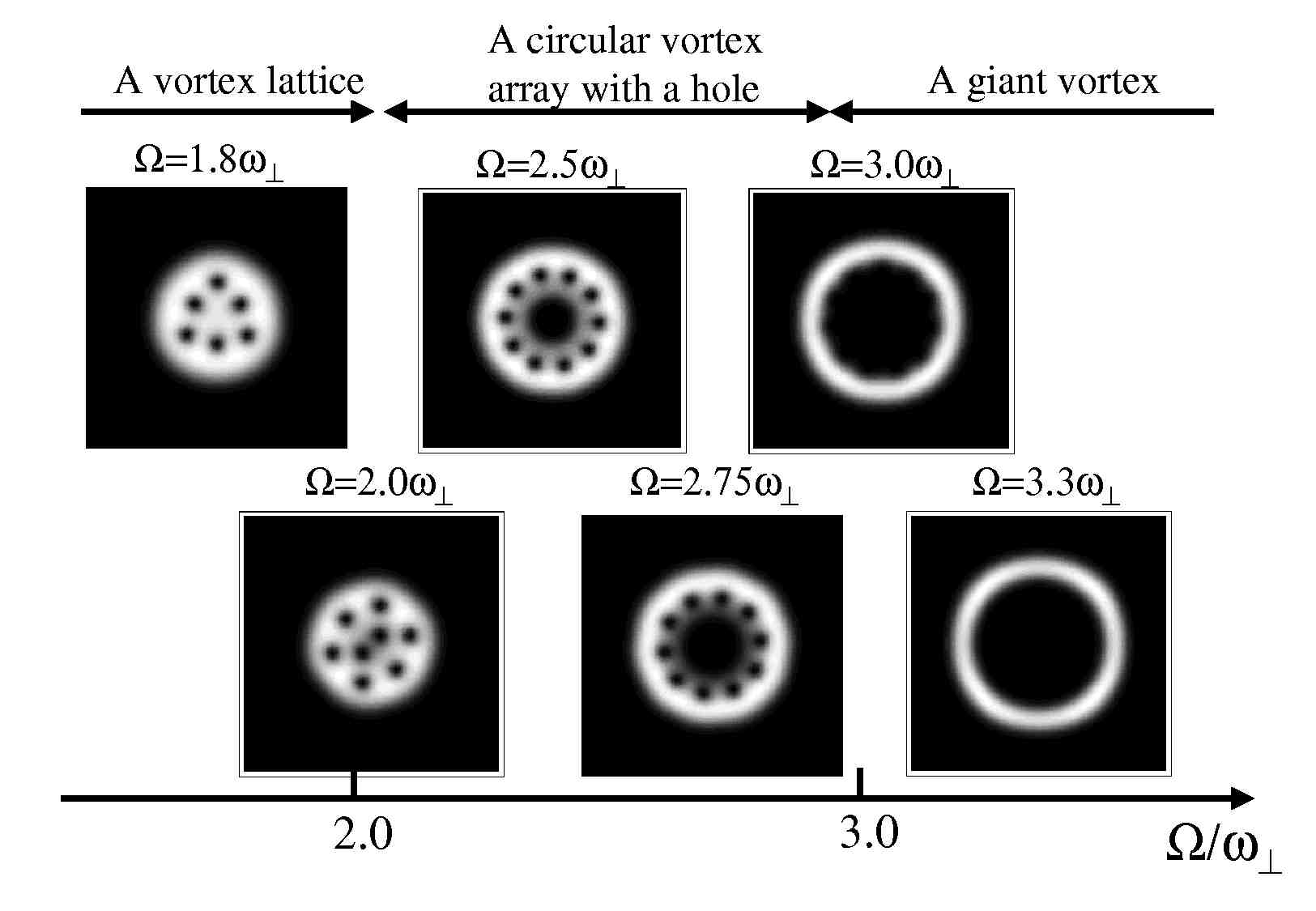}
\caption{The equilibrium vortex states obtained in the numerical simulation for $C=250$.}
\end{figure}
The obtained vortex states for the coupling
coefficient $C=250$ are summarized in Fig. 8 as a function of a rotation
frequency $\Omega$. For small $\Omega$, the equilibrium state is the vortex
lattice state. As $\Omega$ increases, the vortices begin to merge in the
central region. We find that two vortices merge at the center at $\Omega
\simeq 2.2 \omega_{\perp}$. This is the onset of the appearance of a new
vortex state consisting of a density hole and a circular array of vortices.
Further increase of $\Omega$ (above $3.0$) stabilized the giant vortex.
Recently, Kavoulakis and Baym studied the vortex states in a fast rotating
regime under the Thomas-Fermi approximation \cite{Kavoulakis}, which is
effective only for the large interaction energy $C \geq 5000$. It is
difficult to make the numerical simulation for such a strong interaction
because the high rotation will generates an extremely large number of
vortices, whose numerical description would need a very high spatial
resolution. It is interesting to investigate the phase diagram of rich
vortex states in a fast rotating BEC; Kavoulakis and Baym proposed a phase
diagram with triple points between three distinct vortex states in the
interaction strength versus rotation frequency plane \cite{Kavoulakis}.

Experimentally, this giant vortex has been created by Engels {\it et al.}
\cite{Engels}. The authors removed atoms of a rotating condensate from the
central region, which resulted in the further increase of the angular
momentum per particle. Eventually, about 60 vortices merged into a density
hole, which had the long lifetime of many seconds. The combined
harmonic-plus-quartic potential was created in ENS group recently by
superimposing a blue detuned laser with the Gaussian profile \cite{Bretin}.
They observed that, as the rotation frequency approached the frequency of
the harmonic potential, the clear image of a vortex lattice disappeared,
which suggested a transition into a new vortex phase.

\section{CONCLUSIONS}\label{concle}
In this review article, we discussed the dynamics of quantized vortices in
superfluid helium and atomic-gas BECs. Analogous topological defects appear
in almost all fields of physics, including superconductors, magnetic
materials, and cosmology. In each such field, the defects play important
roles in the physics. However, unlike most systems in which the physics of
topological defects usually involve impurities, the vortices in superfluid
helium and atomic BECs exist in very pure systems in which the proper
physics of the topological defects themselves are relevant. This is the
first reason why we are interested in vortices of these systems. The second
reason is that quantum mechanical constraints make the vortices more stable
and definite than those in classical media.

It has been about 40 years since a quantized vortex was discovered in
superfluid helium. As described in this article, research on superfluid
turbulence is about to reach a new stage regarding the similarity between
superfluid turbulence and classical turbulence, a stage in which the nature
of quantized vortices is crucial. In addition, the more recent realization
of atomic-gas BECs have allowed us to visualize and control the dynamics of
vortices, such that the quantum Hall state with a very high density of
vortices may soon be reached \cite{Ho,Cooper,Sinova}. Unconventional
vortices in multicomponent BECs \cite{Kasaken3} is another area that will
likely produce new and interesting results. Given the recent history of
rapid developments in these fields, even more unexpected phenomena are
likely to emerge in the future.

\begin{acknowledgements}
We are grateful for research collaboration with Sergey K. Nemirovskii,
W.F.Vinen, Carlo F. Barenghi and Masahito Ueda. We thank Shigeo Kida, M.E.
Brachet, P.V.E. McClintock, Yasuhide Fukumoto and Tetsuro Nikuni for useful
discussions. M.T. acknowledges W.F.Vinen also for showing the detailed
research history of superfluid turbulence. M.T. is grateful to the Japan
Society for the Promotion of Science for financial support through the
Japan-UK Scientific Cooperative program (Joint Research
project)  and through a Grant-in-Aid for Scientific Research (Grant no.
12640357 and 15340122).
\end{acknowledgements}

\end{document}